\documentstyle[12pt]{article}
\begin{document}
\baselineskip=18pt
\hfill {\large \bf IMSc/98/07/37}

\vfill
\begin{center}
{\Large \bf A New Mechanism for Neutrino Mass}

\vskip 3cm

{\bf P.P.Divakaran} \\
{\it SPIC Mathematical Institute} \\
{\it 92, G.N.Chetty Road, Madras-600 017.}

\vskip .5cm

{\bf and} 

\vskip .5cm

{\bf G.Rajasekaran} \\
{\it Institute of Mathematical Sciences} \\
{\it CIT Campus, Madras-600 113.}
\end{center}

\vskip 2cm

\centerline{\bf Abstract}

\vskip .5cm

A mechanism for generating massive but naturally light Dirac neutrinos is
proposed. It
involves composite Higgs within the standard model as well as some new
interaction beyond the
standard model. According to this scenario, a neutrino mass of 0.1 eV or
higher, signals new physics at energies of 10--100 TeV or lower.
\vfill 

\newpage

The recent announcement of a depletion in the expected number on the earth's
surface of $\nu_\mu's$ originating from cosmic ray interactions in the
atmosphere
[1] has once again focussed attention on the fundamental properties of
neutrinos. The favoured explanation for this effect is that a $\nu_\mu$
oscillates into a neutrino of another family, most likely a $\nu_\tau$,
implying
that at least one neutrino has a nonzero mass. From the reported value of
$\delta
m^2 \approx 10^{-3}$ to $10^{-2} eV^2$, we can conclude that the average
mass of the
two neutrinos involved is bounded approximately by 
$m > \frac{1}{2} \sqrt{\delta m^2} \approx 10^{-1} eV$. 

\vskip .5cm

Massive but light neutrinos have intrigued model-makers for quite some time
now. The most widely discussed possibility is to assume that neutrinos are
Majorana particles, in which case they can be driven to a small mass by
the see-saw
mechanism [2]. If, however, neutrinos turn out to be Dirac particles, we would
require an alternative scenario. In this brief note, we propose such an
alternative. We suggest a simple, qualitative, model-independent line of
reasoning that naturally accommodates light, massive, Dirac neutrinos and draw
from it information regarding the scale at which new physics beyond the
standard
model can be expected to come into play.

\vskip .5cm

Neutrinos are unique in the standard model. They are the only fermions, a part
of
which, namely the right-handed part $\nu_R$, has zero quantum numbers under
$SU(3)
\times SU(2) \times U(1)$ and as a consequence has no gauge interaction. Thus,
if there are no elementary Higgs bosons and if the W,Z, the charged leptons and
the quarks get their masses by dynamical breaking of symmetry induced by the
$SU(3) \times SU(2) \times U(1)$ gauge interactions alone, then neutrinos will
remain massless. In such a case, new interactions going beyond the standard
model
will be required for giving mass to the neutrino. If the mass scale of the new
physics beyond the standard model is large enough, the mass of the neutrino will
remain small.  This would provide a natural mechanism for small neutrino masses.
In contrast, totally arbitrary neutrino masses would result from the
introduction of
elementary Higgs boson, which we discard.
\vskip .5cm

To make our suggestion a little more concrete, let us envisage a picture
in which
the Higgs boson $H$ is a composite of fermions and antifermions bound by the
$SU(3)
\times SU(2) \times U(1)$ gauge forces through some nonperturbative
mechanism [3]. 
In principle, $H$ can be a
combination of  $\bar{t}_L t_R, \bar{b}_L b_R \ldots \bar{d}_L d_R,
\bar{\tau_L}
\tau_R \ldots \bar{e}_L e_R$, but it cannot contain $\bar{\nu}_L \nu_R$ 
of any family, since
$\nu_R$ does not have any gauge interaction. {\it In other words, the effective
Yukawa coupling $H \bar{\nu}_L \nu_R$ vanishes exactly, to all orders in the
$SU
(3) \times SU(2) \times U(1)$ gauge coupling constants}. On the other hand, the
effective Yukawa vertex $H \bar{e}_L e_R$ for the electron (or for any other
charged fermion) exists and it has a form factor characterized by a momentum
scale $\Lambda_H$ which we take to be the electroweak scale $\approx $ 100
GeV as
that is the only relevant scale. We may call the standard model interactions
as 
``allowed'' interactions. In this sense, masses of the charged fermions are
allowed, via the
Yukawa interaction, if $H$ has a nonvanishing vaccuum expectation value, while
neutrino masses are forbidden in the regime of validity of the standard model - the
only way to make neutrinos massive is to invoke forces beyond the standard
model.

\vskip .5cm

We next go to the ``first-forbidden'' approximation, i.e. we include the
nonstandard effects in the lowest nontrivial order. Without being committed to a
specific model, we parametrize the required new physics beyond the standard
model
by effective four-fermion couplings with a Fermi-type couping constant 
generically denoted as $G_X$. The
corresponding mass scale $G^{-1/2}_X$ must be substantially higher than
the mass scale ($\approx$ 100 GeV) of the standard model. This will
generate the first-forbidden coupling $H \bar{\nu}_L \nu_R$ through the
graphs shown in Fig.1, where the shaded vertex is the ``allowed'' Yukawa
vertex with form factor, for a charged fermion which we may take to be a
charged lepton 
$\ell$, so as not to violate $B$ and $L$ at the $X$ vertex. The corresponding 
effective Yukawa coupling constant $f_\nu$ for $\nu$ can be estimated :
$$
f_\nu \approx f_\ell G_X \int^{\Lambda_H} \frac{d^4 p}{\not p \not p} \
\approx \ f_\ell G_X \Lambda^2_H \eqno(1)
$$
where $f_\ell$ is the Yukawa coupling constant for $\ell$ and the
integral is cut off at $\Lambda_H$ because of the form factor of the
composite Higgs. If the Higgs has a nonvanishing vacuum expectation
value, then we get for the neutrino mass 
$$
m_\nu \approx m_\ell G_X \Lambda^2_H \eqno(2)
$$
where $m_\ell$ is the mass of the charged lepton.
For $\Lambda_H \approx 100 GeV$, we arrive at 
$$
G^{-1/2}_X \approx 100 \sqrt{m_\ell /m_\nu} \ GeV\,. \eqno(3)
$$
A lower bround on $m_\nu$ thus results in an upper bound on the
scale of new physics $G_X^{-1/2}$. Also,
the lower the mass of $\ell$ to which $\nu$ couples at the $\bar{\ell}_R
\nu_R X$ vertex, the lower is the bound on $G^{-1/2}_X$ and the best
bound is obtained for the charged lepton of lowest mass $\ell$ to which
the neutrino of mass $m_\nu$ couples.

If the dominant mixing of the atmospheric $\nu_\mu$ is with $\nu_\tau$,
the best bound on $G^{-1/2}_X$ is realised for $\ell = \mu$ in the above
formula :
$$
G^{-1/2}_X \leq 10^5 - 10^6 GeV. \eqno(4)
$$
If the massive neutrino couples also to $e_R$ (for which there is no clear
evidence), this bound will be reduced by a factor of about 10. In
contrast to the see-saw mechanism where $m_\nu$ depends linearly on the
mass of the heavy right-handed Majorana neutrino, in our formula (2),
$m_\nu$ depends on the square of the mass-scale of new physics. As a
result, in the scenario envisaged here, new physics would occur at much
lower energies than with the see-saw mechanism and so our proposal can
be confronted with experiment much earlier and either confirmed or ruled
out.

We discuss briefly two illustrative possibilities of new physics 
beyond the standard model (SM), that would lead to the two
types of effective four-fermion couplings introduced in Fig.1.  We must
note that the
type (a) coupling (Fig.1(a)) is consistent with SM symmetry, but type (b)
coupling
(Fig.1(b)) violates $SU(2) \times U(1)$. Type (a) can be obtained by the
exchange of
a charged or neutral scalar boson $S$ (as shown in Fig.(2)), with
coupling constant
$h$ and mass $m_S \gg$ 100 GeV. In this case, $G_X$  can be replaced by
$$
G_X \approx \frac{h^2}{m_S^2}\,, \eqno(5)
$$
and hence the upper bound on $m_S$ would be smaller than $10^5-10^6$ GeV,
if $h$ is less
than unity. If all elementary scalars are forbidden (which is not necessary
for our
argument on the neutrino mass), these $S$ bosons also could be composite,
but formed
by forces beyond the standard model. (For obvious reasons $S^o$ must have zero
vacuum expectation value).

The SM-symmetry violating four-fermion coupling (type (b)) occurs in a
large class
of models in which the $W$ boson of the SM mixes with a heavier $W$ boson that
couples to righthanded fermions. The best model of this kind is the one in which
$SU(2)_L \times U(1)$ of the SM is extended [4] to $SU(2)_L \times SU(2)_R
\times
U(1)$. This has two pairs of charged $W$ bosons, $W^\pm_L$ and $W^\pm_R$.
The mass
eigenstates $W^\pm_1$ and $W^\pm_2$ can be expressed through the mixing angle
$\zeta$ :
$$  
W_1 = W_L \cos \zeta + W_R \sin \zeta \eqno(6)
$$
$$
W_2 =  -W_L \sin \zeta + W_R \cos \zeta \,. \eqno(7)
$$
One identifies $W_1$ with the known $W$ boson and $W_2$ is presumed to be heavy. The
current experimental limits are [5]
$$
|\zeta| < 10^{-2} - 10^{-3}  \eqno(8)
$$
$$
\beta \equiv \frac{m_{W_1}^2}{m_{W_2}^2} \ < 0.02. \eqno(9)
$$
We also have a theoretical bound [6]
$$
|\zeta| < \beta \,. \eqno(10)
$$

Fig.3 shows the $W$-exchange graphs that generate the required coupling of
Fig.1b.
The effective Fermi-coupling constant arising from the sum of these two
graphs can be estimated to be
$$
G_X \approx  g^2 \cos \zeta \sin \zeta \pmatrix{\frac{1}{m_{W_1}^2} -
\frac{1}{m_{W_2}^2}} \eqno(10)
$$
$$
\approx \left(\frac{\zeta}{\beta}\right) \frac{g^2}{m_{W_2}^2} \
\leq \ \frac{g^2}{m_{W_2}^2} \,. \eqno(11)
$$
Combining (4) and (11) and using the value of the $SU(2)_L$ gauge coupling
constant
$g$, we get the bound
$$
m_{W_2} \leq 10^4 - 10^5 GeV. \eqno(12)
$$

Our conclusions are : (i) The physics of a composite Higgs or, more generally,
dynamical symmetry breaking within the standard model followed by new
interactions
beyond the standard model provides a natural mechanism for generating very small
neutrino masses. (ii) Within this scenario, a finite but small Dirac
mass for neutrinos may be regarded as a signal that interesting new physics
can be
expected at an energy scale of 10--100 TeV or lower.

\vskip .5cm

\noindent {\bf Acknowledgements} 

\vskip .5cm

We thank Ramesh Anishetty, Anjan Joshipura, Sandip Pakvasa and Xerxes Tata for
discussions and criticisms. PPD would like to acknowledge the use of the
facilities of the Institute of Mathematical
Sciences, Madras and the Tata Institute of Fundamental Research,
Bombay.

\newpage

\noindent {\bf References and Footnotes}

\begin{enumerate}
\item Y.Fukuda {\it et al.} (Super-Kamiokande Collaboration), hep-ex/9803006 and
hep-ex/ \linebreak 9805006 ; T.Kajita, Talk at {\it `Neutrino 98'},
Takayama, Japan.

\item M.Gell-Mann, P.Ramond and R.Slansky, in {\it Supergravity} (Ed. P. Van
Nieuwenhuizen and D.Z.Freedman, North Holland, Amsterdam, 1979), p.315 ;
T.Yanagida, in {\it Proc. of the Workshop on the Unified Theory and Baryon
Number in
the Universe} (Ed. O.Sawada and A.Sugamoto) KEK Report No.79-18, Tsukuba, Japan,
1979.

\item This nonperturbative mechanism may have its origin in the as-yet-unsolved
problem of infra-red divergences that afflict the unbroken phase of the
nonabelian
gauge theory.

\item J.C.Pati and A.Salam, Phys. Rev. {\bf D10}, 275 (1974) ;
R.N.Mohapatra and J.C.Pati, {\it Ibid.} {\bf 11}, 566 (1975) ; {\bf 11}, 2558
(1975) ; R.N.Mohapatra and G.Senjanovic, {\it Ibid.} {\bf 12}, 1502, (1975) ;
P.Langacker and S.Uma Sankar, {\it Ibid.} {\bf 40}, 1569 (1989).

\item Review of Particle Physics : C.Caso {\it et al.} (Particle Data
Group), European Physical Journal {\bf C3}, 1 (1998).

\item E.Masso, Phys. Rev. Lett. {\bf 52}, 1956 (1984).

\end{enumerate}

\newpage

\noindent {\bf Figure Captions}

\begin{description}
\item[Fig.1.] Generation of the Yukawa coupling of the neutrino, from
that of the charged lepton with form factor denoted by the shaded
vertex, through new physics represented by the four-fermion coupling of
two types (a) and (b). (To each diagram, one must add a corresponding
diagram with
{\it all} fermion lines reversed).

\item[Fig.2.] Type (a) coupling illustrated by scalar boson exchanges.
\item[Fig.3.] Type (b) coupling illustrated by exchanges of $W_1$ and
$W_2$ gauge bosons which are mixtures of $W_L$ and $W_R$.
\end{description}

\newpage
\begin{figure}
\vskip 6 truecm
\includegraphics{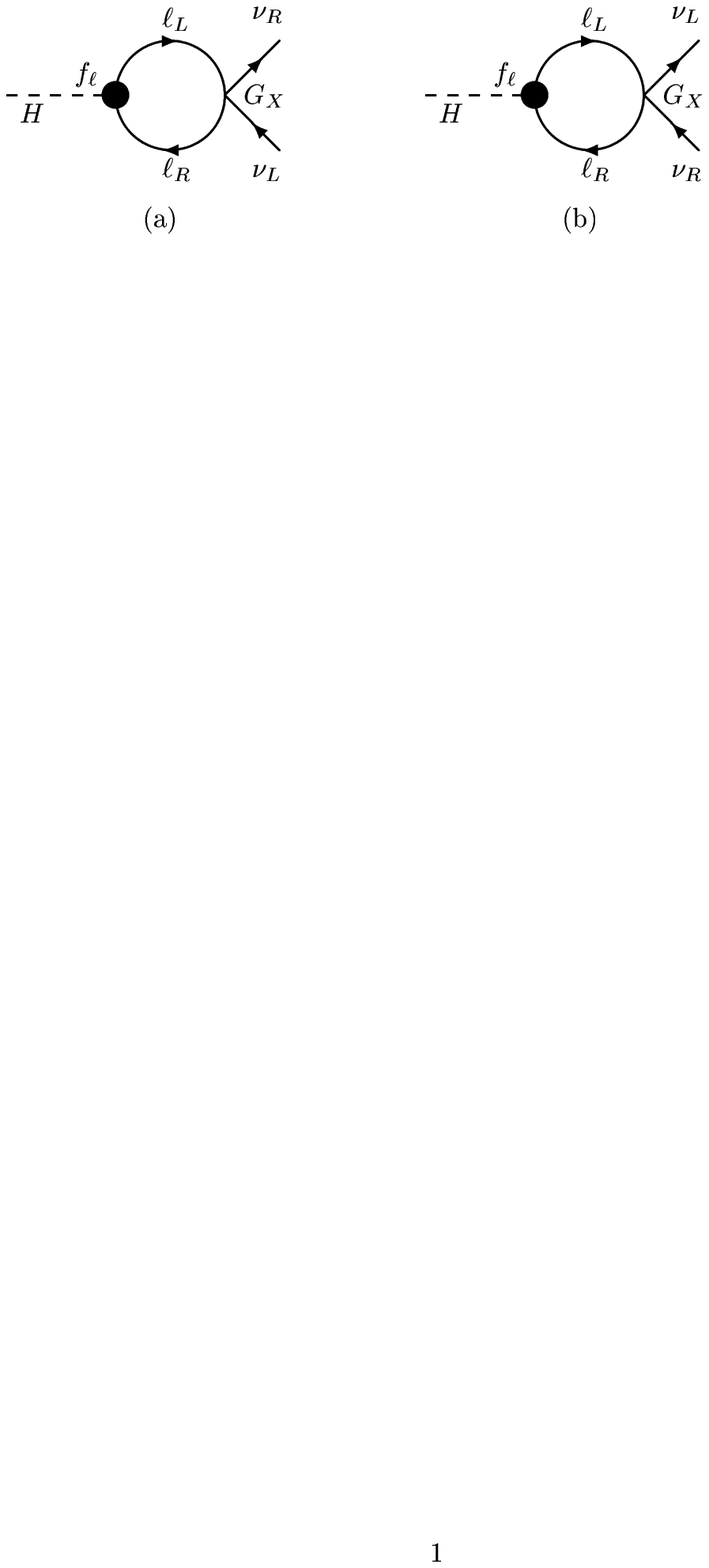}

\caption{}
\end{figure}

\begin{figure}
\vskip 6 truecm
\includegraphics{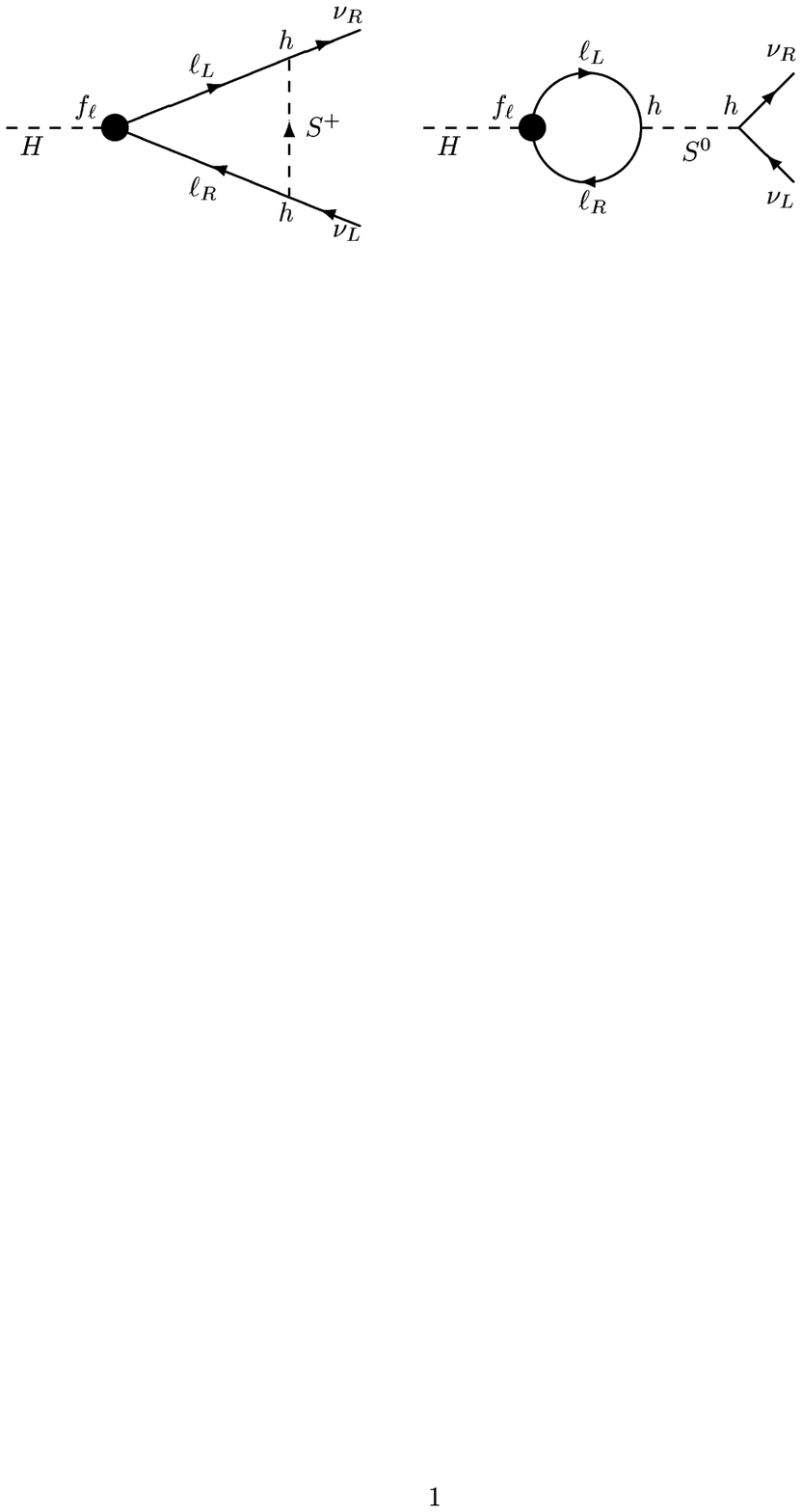}

\caption{}
\end{figure}

\begin{figure}
\vskip 6 truecm
\includegraphics{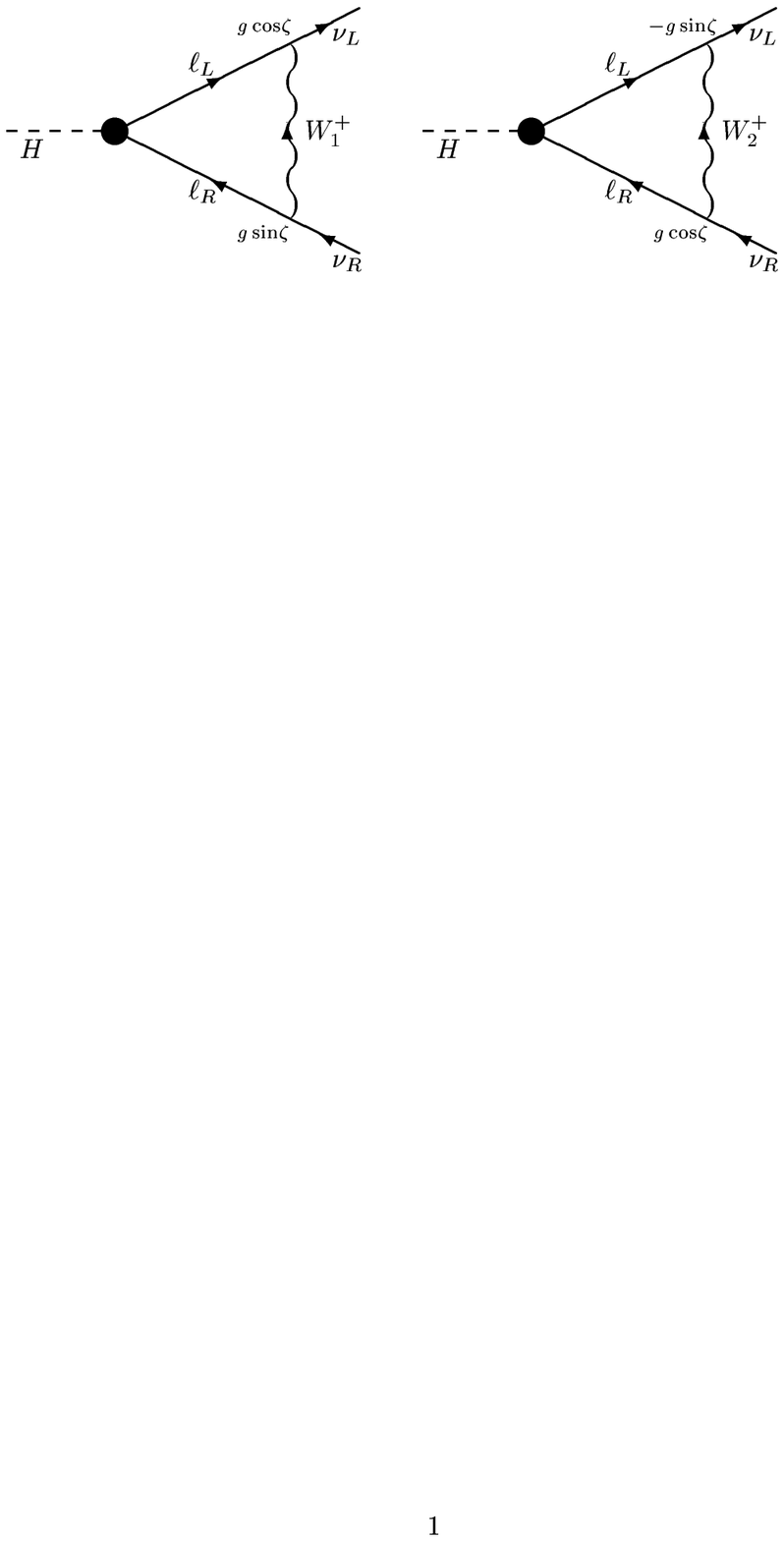}

\caption{}
\end{figure}

\end{document}